\documentclass[12pt,preprint]{aastex}

\usepackage{epsfig}
\usepackage{lscape}

\received{01 November 2004}
\accepted{17 December 2004}

\newcommand{\dechms}[4]{$#1^{\rm h}#2^{\rm m}#3\mbox{$^{\rm s}\mskip-7.6mu.\,$}#4$}

\newcommand{\decdms}[4]{$#1^{\circ}#2'#3\mbox{$''\mskip-7.6mu.\,$}#4$}
\newcommand{\msec}[2]{$#1\mbox{$''\mskip-7.6mu.\,$}#2$}
\newcommand{\mmsec}[2]{$#1\mbox{$^s\mskip-7.6mu.\,$}#2$}
\newcommand{\mdeg}[2]{$#1\mbox{$^\circ \mskip-7.6mu.\,$}#2$}
\newcommand{\mtsec}[2]{$#1\mbox{$^{\rm s}\mskip-7.6mu .\,$}#2$}

\newcommand{\Rsun}{R$_{\odot}$}

\begin{document}

\title{Multi-epoch VLBA observations of T Tauri South}

\author{Laurent Loinard\footnote{Centro de Radiostronom\'{\i}a y
    Astrof\'{\i}sica, Universidad Nacional Aut\'onoma de M\'exico,
    Apartado Postal 72--3 (Xangari), 58089 Morelia, Michoac\'an,
    M\'exico; l.loinard@astrosmo.unam.mx}, Amy J.\ 
  Mioduszewski\footnote{ National Radio Astronomy Observatory, Array
    Operations Center, 1003 Lopezville Road, Socorro, NM 87801, USA},
  Luis F.\ Rodr\'{\i}guez\altaffilmark{1},\\
  Rosa A.\ Gonz\'alez\altaffilmark{1}, M\'onica I.\ 
  Rodr\'{\i}guez\altaffilmark{1}, and Rosa M.\ Torres\altaffilmark{1}}

\begin{abstract}
  In this {\it Letter}, we present a series of seven observations of
  the compact, non-thermal radio source associated with T Tauri South
  made with the Very Long Baseline Array over the course of one year.
  The emission is found to be composed of a compact structure most
  certainly originating from the magnetosphere of an underlying
  pre-main sequence star, and a low-brightness extension which may
  result from reconnection flares at the star-disk interface. The
  accuracy of the absolute astrometry offered by the VLBA allows very
  precise determinations of the trigonometric parallax and proper
  motion of T Tau South. The proper motion derived from our VLBA
  observations agrees with that measured with the VLA over a similar
  period to better than 2 mas yr$^{-1}$, and is fully consistent with
  the infrared proper motion of T Tau Sb, the pre-main sequence M star
  with which the radio source has traditionally been associated. The
  parallax, $\pi$ = 7.07 $\pm$ 0.14 mas, corresponds to a distance of
  141.5$^{+2.8}_{-2.7}$ pc.
\end{abstract}


\section{Introduction}

T Tauri was initially identified as a single optical star, with
unusual variability and peculiar emission lines (Barnard 1895, Joy
1945 and references therein). Early infrared observations then
revealed the existence of a heavily obscured companion (hereafter T
Tau S) located about \msec{0}{7} to the south of the visible star
(Dyck, Simon \& Zuckerman 1982), and most likely gravitationally bound
to it (Ghez et al.\ 1991).  Recently, this infrared companion was
itself found to contain two sources (T Tau Sa and T Tau Sb --Koresko
2000; K\"ohler, Kasper \& Herbst 2000) in rapid relative motion
(Duch\^ene, Ghez \& McCabe 2002; Furlan et al.\ 2003). Thus, T Tau is
now acknowledged to be at least a triple stellar system. At radio
wavelengths, T Tau has long been known to be a double source
(Schwartz, Simon \& Campbell 1986).  The northern radio component is
associated with the optical star, and mostly traces the base of its
thermal jet (e.g.\ Johnston et al.\ 2003), whereas the southern radio
source is related to the infrared companion, and is thought to be the
superposition of a compact component of magnetic origin and an
extended halo, presumably related to stellar winds (Johnston et al.\ 
2003, Loinard et al.\ 2003).

The relative motions between the various components of the T Tau
system have recently been under enhanced scrutiny, following the
suggestion, based on multi-epoch VLA observations, that one of the
components might have seen its path dramatically altered by a recent
chaotic encounter (Loinard et al.\ 2003). This interpretation was
disputed by Johnston et al.\ (2003, 2004b) and Tamazian (2004), who
fitted the same VLA data with stable orbits. It is noteworthy,
however, that the residuals between their best fits and the actual
radio positions (\msec{0}{03} -- \msec{0}{04}) are often significantly
larger than the nominal observational errors ($\lesssim$
\msec{0}{01}), especially at recent epochs (see Tab.\ 6 in Johnston et
al.\ 2004b).  To resolve this discrepancy, Johnston et al.\ (2004a,
2004b) proposed that the structure of T Tau S at radio wavelengths was
affected by erratic internal variations, which made the centroid of
the VLA source dither around the position of the underlying PMS star.
To reconcile the radio observations obtained in the last few years
with the orbital fits, the VLA source centroid needs to have moved
about 25 mas yr$^{-1}$ faster than the associated PMS star. Clearly,
this would render the existing 20 years of VLA observations useless as
tracers of the stellar trajectories, in spite of the high quality
astrometry naturally provided by radio interferometry.

\begin{table*}[!ht]
\caption{Observational parameters.}
\centering
\begin{tabular}{cccccc}
\hline
\\[-0.25cm]
UT Date    & Epoch     & Synthesized Beam & r.m.s. & Flux density & Peak flux\\%
           & (yr)      & $\Delta \theta_{max}$ $\times$ $\Delta \theta_{min}$; P.A. & ($\mu$Jy beam$^{-1}$) & (mJy) & (mJy beam$^{-1}$)\\%
           &           & (mas $\times$ mas; degrees) \\%
\\[-0.25cm]
\hline
\\[-0.25cm]
24.09.2003 & 2003.7300 & 1.88 $\times$ 0.78; ~$-$8.99  &  70.5  &  1.69 & 1.43 \\%
18.11.2003 & 2003.8804 & 1.87 $\times$ 0.79; ~$-$5.19  &  64.3  &  1.70 & 1.17 \\%
15.01.2004 & 2004.0387 & 2.17 $\times$ 1.00; $+$11.94  &  68.0  &  1.03 & 0.84 \\%
26.03.2004 & 2004.2349 & 1.90 $\times$ 0.80; ~$-$2.24  &  65.5  &  1.51 & 0.76 \\%
13.05.2004 & 2004.3657 & 1.92 $\times$ 0.76; $-$10.66  &  76.3  &  1.79 & 1.18 \\%
08.07.2004 & 2004.5183 & 1.90 $\times$ 0.78; ~$-$6.00  &  62.9  &  1.13 & 0.54 \\%
16.09.2004 & 2004.7090 & 2.06 $\times$ 0.80; ~$-$8.94  &  68.7  &  1.70 & 0.98 \\%
\\[-0.25cm]
\hline
\end{tabular}
\end{table*}

The non-thermal mechanisms at the origin of the compact radio emission
in T Tau S require the presence of an underlying, magnetically active
star (Skinner 1993). Specifically, the emission is expected to be
either gyrosynchrotron radiation associated with reconnection flares
in the stellar magnetosphere and at the star-disk interface; or
coherently amplified cyclotron emission from magnetized accretion
funnels connecting the disk to the star (Dulk 1985, Feigelson \&
Montmerle 1999, Smith et al.\ 2003).  In all cases, the emission is
produced within less than about 10 stellar radii (roughly 30 \Rsun) of
the star itself. Indeed, 3.6 cm VLBI observations recently revealed
the existence, near the expected position of T Tau Sb, of a source
with an angular size less than about 15 \Rsun\ (Smith et al.\ 2003).
Because it is so small, any structural changes in this compact radio
component would occur on such small scales that the effects on the
astrometry would be very limited. Thus, observations focusing on it
should accurately trace the path of the underlying PMS star.

\section{Observations}

Here, we present the results of a series of seven continuum 3.6 cm
(8.42 GHz) observations of T Tau S obtained every two months between
September 2003 and September 2004 with the 10-element Very Long
Baseline Array (VLBA) of the National Radio Astronomy Observatory
(NRAO; Tab.\ 1).  Since the VLBA is only sensitive to compact emission
structures with high surface brightness, it will effectively filter
out the extended radio halo of T Tau S, and provide images of the
non-thermal source alone.  Our phase center was at $\alpha_{J2000.0}$
= \dechms{04}{21}{59}{4263}, $\delta_{J2000.0}$ =
+\decdms{19}{32}{05}{730}, the position of the compact source detected
by Smith et al.\ (1999). Each observation consisted of a series of
cycles with two minutes spent on source, and one minute spent on the
phase-referencing quasar J0428+1732 ($\alpha_{J2000.0}$ =
\dechms{04}{28}{35}{633679}, $\delta_{J2000.0}$ =
+\decdms{17}{32}{05}{58799}), located \mdeg{2}{6} away.  The secondary
quasar J0431+1731 ($\alpha_{J2000.0}$ = \dechms{04}{31}{57}{379244},
$\delta_{J2000.0}$ = +\decdms{17}{31}{35}{77538}) was also observed
periodically (about every 30 minutes); it was not used in the phase
calibration process, but served as a check on the final astrometry.
Based on the dispersion on the measured position of that quasar, we
estimate our astrometric uncertainties to be about 0.25 mas in both
$\alpha$ and $\delta$. Data editing, amplitude calibration, and fringe
fitting (carried out only on the calibrators, since the target is much
too weak) were made in a standard way using NRAO's AIPS software.
Once calibrated, the visibilities were imaged with a pixel size of
0.25 mas after weights intermediate between natural and uniform were
applied.

\begin{figure*}
\centerline{\includegraphics[height=0.90\textwidth,angle=-90]{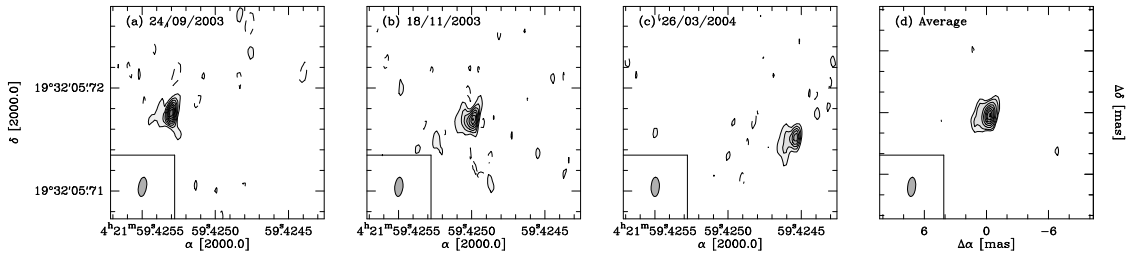}}
\caption{(a)--(c): 3.6 cm images of the T Tau system obtained at three 
  different epochs. The first contour and the contour interval are at
  0.165 mJy beam$^{-1}$. (d) Average of the 6 epochs. The first
  contour is at 0.08 mJy beam$^{-1}$ and the subsequent ones increase
  exponentially by a factor $\sqrt{2}$. The synthesized beam is
  indicated at the bottom left corner of each panel.}
\end{figure*}

\section{Structure of the emission}

The compact radio source associated with T Tau S was detected at all
seven epochs with high signal-to-noise ratio (Tab.\ 1). Its flux
density varies significantly, from a maximum of nearly 1.8 mJy, down
to a minimum of just over 1.0 mJy.  Moreover, its peak brightness in
mJy beam$^{-1}$ is found to be always significantly smaller than its
flux density in mJy (on average by 35 \% --Tab.\ 1).  This suggests
that the emission is somewhat extended since equal numerical values
are expected for a point source.  This possibility gains further
credibility when one considers the restored images (Fig.\ 1a-c), where
the emission is consistently found to be composed of a compact source
(most certainly associated with a stellar magnetosphere) and a low
brightness ``spur'' extending eastwards.  Although this latter
component was not reported by Smith et al.\ (2003), inspection of
their published image (their Fig.\ 1) reveals a source structure quite
similar to that found here, with a weak spur extending towards the
north-east of a compact component.  When our seven observations are
combined, the low brightness spur becomes even more evident (Fig.\ 
1d); its position angle is between +70$^{\circ}$ and +80$^{\circ}$,
and its deconvolved extent in that direction is about 1 mas. Since the
measured parallactic and proper motions (Sect.\ 3.2) imply a maximum
displacement of about 0.02 mas during any one of our 6-hour observing
runs, the measured size is not affected by smearing effects. The
observed spur is also very unlikely to be a consequence of random
errors in the phase calibration, since its characteristics repeat
themselves from one epoch to the next.

T Tau S contains two infrared sources, so one might be tempted to
associate the eastern spur to this preferred direction of the system.
However, extrapolating from the latest published infrared observations
(Furlan et al.\ 2003; Duch\^ene et al., in prep.), we estimate that
the position angle between T Tau Sb and T Tau Sa at the median epoch
of our VLBA observations must have been about 110$^{\circ}$,
significantly larger than the position angle of the spur. Another
unlikely possibility is that it would trace a weak secondary
component. Given the short distance which would then separate the two
VLBA sources, orbital motions should have been easily detected over
the course of our observations.  Interestingly, the spur is nearly
exactly perpendicular to the jet acknowledged to be powered by T Tau S
(Solf \& B\"ohm 1999). This might suggest that the low brightness
extension is associated with an accretion disk, and results from
reconnection flares at the star-disk interface. Indeed, Duch\^ene et
al. (2002) reported the detection of a wavelength-dependent infrared
excess around T Tau Sb, which strongly suggests the existence of an
accretion disk there. Typically, the inner edge of that disk is
expected to be at a radius of about $5R_*$, where $R_*$ is the stellar
radius (Ostriker \& Shu 1995). A low-mass PMS star such as T Tau Sb is
expected to have a radius of about 3 \Rsun\ (Siess, Dufour \&
Forestini 2000).  Hence, the inner disk edge is expected to be at 15
\Rsun\ (0.07 AU). At a distance of 141.5 pc (see Sect.\ 3.2), this
corresponds to an angular scale of about 0.5 mas. After convolution
with our beam size in $\alpha$ (on average 0.82 mas), we obtain an
expected observed extent of 0.9 mas for the spur. This is in very good
agreement with the measured extent. In this scheme, it is, however,
somewhat puzzling that we only see structure on one side of the star.

\section{Astrometry}

As Figs.\ 1a to 1c readily show, the absolute position of the source
changes significantly from one epoch to the next. These displacements
most certainly result from parallactic and proper motions\footnote{At
  least one more year of observations will be needed before
  acceleration terms can be measured reliably.}, and can be modeled in
terms of the source position at epoch J2000.0 ($\alpha_{J2000.0}$ and
$\delta_{J2000.0}$), its proper motion in right ascension and
declination ($\mu_\alpha$ and $\mu_\delta$), and its parallax $\pi$.
These five astrometric elements were deduced from the measured
positions of the source by minimizing the $\chi^2$ associated with
this description using an iterative scheme (Fig.\ 2). The results for
the five astrometric parameters are:

\begin{center}
\begin{tabular}{rl}  
\hline
$\alpha_{J2000.0}$ =       & \dechms{04}{21}{59}{424015} $\pm$ \mtsec{0}{000009} \\%
$\delta_{J2000.0}$ =       & \decdms{19}{32}{05}{71957} $\pm$ \msec{0}{00013} \\%
$\mu_\alpha \cos \delta$ = & 3.29 $\pm$ 0.30 mas yr$^{-1}$ \\%
$\mu_\delta$ =             & $-$0.75 $\pm$ 0.31 mas yr$^{-1}$ \\%
$\pi$ =                    & 7.07 $\pm$ 0.14 mas \\%
\hline
\end{tabular}
\end{center}

\smallskip

\noindent
The post-fit r.m.s.\ is found to be 0.25 mas in both $\alpha$ and
$\delta$, in agreement with our expected individual positional errors.

\begin{figure*}
\centerline{\includegraphics[height=0.74\textwidth,angle=-90]{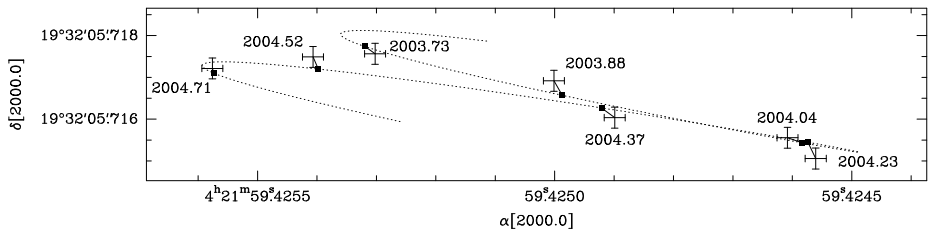}}
\caption{Trajectory on the plane of the sky of the compact radio 
  component in T Tau S. The crosses represent the measured positions of
  the source with their error bars, the dotted line shows the best fit to
  the data, and the black squares display the positions of the source as
  predicted by the best fit for each observed epoch.}
\end{figure*}

The parallax reported here for T Tau corresponds to a distance $d$ =
141.5$^{+2.8}_{-2.7}$ pc, a value much more precise than, and barely
within the 1$\sigma$ errorbar of that obtained by the Hipparcos
satellite for T Tau N ($d$ = 177$^{+70}_{-39}$ pc; Perryman et al.\ 
1997). With this measurement, T Tau becomes the second PMS star in
Taurus with a distance known to this level of precision.  The first
was V773 Tau, for which Lestrade et al.\ (1999) determined a parallax
of 6.74 $\pm$ 0.25 mas ($d$ = 148.3$^{+5.7}_{-5.3}$ pc) using
multi-epoch global VLBI observations.  For V773 Tau, the VLBI distance
was also significantly more precise than, and only marginally within
the 1$\sigma$ errorbar of the Hipparcos value ($d$ = 101$^{+40}_{-22}$
pc). As already discussed by Bertout et al.\ (1999), one must exercise
caution when using Hipparcos parallaxes for faint individual PMS
stars. In particular, while Hipparcos suggested that the southern part
of the Taurus complex might be somewhat farther than its central
regions (170 pc vs.\ 125 pc; Bertout et al.\ 1999), we find T Tau
(which is part of Taurus South) and V773 Tau (which belongs to the
central parts of Taurus) to be at very similar distance.  Indeed, the
two VLBI measurements are consistent with the entire Taurus complex
lying at 145 $\pm$ 4 pc.

\begin{table}
\caption{VLA source positions.}
\centering
\begin{tabular}{lcc}
\hline
Epoch & $\alpha_{2000}$ (04$^h$ 21$^m$) & $\delta_{2000}$ (19$^\circ$ 32$'$) \\%
\\[-0.3cm]
2001.052 & \mmsec{59}{42502} $\pm$ \mmsec{0}{00007} & \msec{05}{7179} $\pm$ \msec{0}{0009} \\%
2003.636 & \mmsec{59}{42574} $\pm$ \mmsec{0}{00006} & \msec{05}{7191} $\pm$ \msec{0}{0008} \\%
\hline
\end{tabular}
\end{table}

To decide if the magnetically active PMS star responsible for the
compact radio source detected here can be identified with the infrared
source T Tau Sb --as is usually believed; and if the VLA observations
can be used to trace its motion, one would ideally like to compare the
absolute positions of the infrared, VLA, and VLBA sources at a common
time. The fit presented in Sect.\ 3.2 can provide accurate estimates
of the absolute position of the VLBA source at any time during, or
within a few years of our observing timespan.  Unfortunately, this
cannot easily be compared with the infrared and VLA positions, because
the infrared data lack accurate absolute astrometry information, while
the VLA and VLBA reference frames may not match perfectly in spite of
being both based on distant quasars.  The latter effect is a
consequence of the differing UV plane coverages of the VLA and the
VLBA, which will tend to make them sensitive to different components
of even the same quasar.  Consequently, it is preferable to compare
the motions of the sources rather than their positions; if two sources
share the same proper motion, it is very unlikely that they are
different objects.

The proper motion of the VLA source was measured over a timespan
similar to that covered by our VLBA observations using the two most
recent 2 cm images available to us: the January 19, 2001 observation
reported in Loinard et al.\ (2003) and Johnston et al.\ (2003), and a
more recent (still unpublished) image we obtained on August 19, 2003.
The latter was processed following the exact same procedure as that
used in Loinard et al.\ (2003). The phase-referencing quasar was
0403+260 ($\alpha_{J2000.0}$ = \dechms{04}{03}{05}{5860},
$\delta_{J2000.0}$ = +\decdms{26}{00}{01}{502}) for both observations.
The calibrated visibilities were restored using pixels of \msec{0}{1}
and weights intermediate between uniform and natural. The VLA source
positions (Tab.\ 2) were determined from 2D Gaussian fits, and were
corrected for parallactic motions. They yield proper motions of:

\begin{center}
\begin{tabular}{rl}  
$\mu_\alpha \cos \delta$ = & 3.94 $\pm$ 0.50 mas yr$^{-1}$ \\%
$\mu_\delta$ =             & 0.46 $\pm$ 0.47 mas yr$^{-1}$, \\%
\end{tabular}
\end{center}

\noindent
very similar to those found for the VLBA source. The differences ($<$
2 mas yr$^{-1}$) are at least an order of magnitude smaller than the
values required to reconcile the VLA observations with the orbital
fits proposed by Johnston et al.\ (2004b). Instead, it appears that
the VLA and the VLBA observations trace the trajectory of the same
magnetically active star.

Because of the lack of absolute astrometry information for the
infrared observations, the proper motions required to compare the radio
and infrared data must be measured relatively to a third source. Here,
we shall register all motions on T Tau N, which is seen both at radio
and infrared wavelengths, and has a well-determined linear proper
motion (e.g.\ Loinard et al.\ 2003). For the radio source, we obtain
a relative motion between T Tau S and T Tau N of:

\begin{center}
\begin{tabular}{rlcrl}  
$\mu_\alpha \cos \delta$ = & $-$8.3 $\pm$ 0.8 & ~~~ or ~~~~ & $-$8.9 $\pm$ 0.7 mas yr$^{-1}$~ \\%
$\mu_\delta$ =             & +13.2 $\pm$ 0.8   &             & +12.0 $\pm$ 0.7 mas yr$^{-1}$, \\%
\end{tabular}
\end{center}

\noindent
depending on whether the VLA or the VLBA data are used. Using the
infrared data in Duch\^ene et al.\ (in prep.), we can also estimate
the relative motion between the infrared source T Tau Sb and T Tau N
during a timespan (November 2001 -- December 2003) similar to that
covered by the radio observations. We obtain:

\begin{center}
\begin{tabular}{rl}  
$\mu_\alpha \cos \delta$ = & $-$10.4 $\pm$ 3.5 mas yr$^{-1}$ \\%
$\mu_\delta$ =             & 12.4 $\pm$ 3.5 mas yr$^{-1}$. \\%
\end{tabular}
\end{center}

\noindent
The excellent agreement between the infrared and the radio proper
motions confirms that the magnetically active PMS star generating the
compact radio emission is T Tau Sb, and that the radio data (from both
VLA and VLBA) can be used to trace its trajectory.

\section{Conclusions and perspectives}

The VLBA data presented here have allowed a detailed study of the
structure and astrometry of the compact radio source associated with T
Tau S. The emission is found to be composed of a compact core most
certainly originating from a stellar magnetosphere, and a
low-brightness extension which may result from reconnection flares at
the star-disk interface. The accuracy of the absolute astrometry
offered by the VLBA has made it possible for us to demonstrate that
the infrared, the VLA, and the VLBA observations all trace the same
underlying star (T Tau Sb). This implies, in particular, that any
orbital fit to the T Tau system must be able to reproduce the
positions measured by the VLA in the last 20 years. Finally, the VLBA
data have allowed us to measure the distance to T Tau with
unprecedented accuracy: $d$ = 141.5$^{+2.8}_{-2.7}$ pc. This is in
good agreement with the traditionally accepted value for Taurus ($d$ =
140 $\pm$ 10 pc --Kenyon, Dobrzycka \& Hartmann 1994), and seemingly
rules out the possibility of large distance gradients across Taurus.

\begin{acknowledgements}
  We are indebted to Walter Brisken and Gaspard Duch\^ene who provided
  VLBA pulsar data with which to checked our astrometry results, and
  infrared positions ahead of publication, respectively. L.L., L.F.R.,
  and R.A.G acknowledge the financial support of DGAPA, UNAM and
  CONACyT, M\'exico; A.J.M.  acknowledges support from the NRAO.  The
  National Radio Astronomy Observatory is a facility of the National
  Science Foundation operated under cooperative agreement by
  Associated Universities, Inc.
\end{acknowledgements}

\end{document}